%% file: DroidLink-main.tex
\documentclass{article}

\usepackage{geometry}
\usepackage{multicol,graphicx,amssymb,mathrsfs,amsmath,pifont,amscd,latexsym,amsthm,color,fancyhdr,CJK,url,hyperref,longtable, pifont, subfigure,epstopdf,setspace,multirow,colortbl,tabularx,threeparttable, booktabs, verbatim, bbm,listings,balance,xcolor}
\usepackage[linesnumbered,boxed,algosection]{algorithm2e}

\newtheorem{Definition}{Definition}
\newcommand{\tabincell}[2]{\begin{tabular}{@{}#1@{}}#2\end{tabular}}

\begin{document}
\lstset{
basicstyle=\ttfamily,
columns=fullflexible,
showstringspaces=false,
keywordstyle= \color{ blue!70},commentstyle=\color{red!50!green!50!blue!50},
frame=shadowbox,
rulesepcolor= \color{ red!20!green!20!blue!20}
}

\title{DroidLink: Automated Generation of Deep Links for Android Apps\footnote{Corresponding: liuxuanzhe@pku.edu.cn, mayun@pku.edu.cn}}
\author{Yun Ma, Xuanzhe Liu, Ruogu Du, Ziniu Hu, Yi Liu, Meihua Yu, Gang Huang}

\maketitle

\begin{abstract}
The mobile application (app) has become the main entrance to access the Internet on handheld devices. Unlike the Web where each webpage has a global URL to reach directly, a specific ``content page" of an app can be opened only by exploring the app with several operations from the landing page. The interoperability between apps is quite fixed and thus limits the value-added ``linked data" between apps. Recently, deep link has been proposed to enable targeting and opening a specific page of an app externally with an accessible uniform resource identifier (URI). However, implementing deep link for mobile apps requires a lot of manual efforts by app developers, which can be very error-prone and time-consuming. In this paper, we propose \emph{DroidLink} to automatically generating deep links for existing Android apps. We design a deep link model suitable for automatic generation. Then we explore the transition of pages and build a navigation graph based on static and dynamic analysis of Android apps. Next, we realize an updating mechanism that keeps on revisiting the target app and discover new pages, and thus generates deep links for every single page of the app. Finally, we repackage the app with deep link supports, but requires no additional deployment requirements. We generate deep links for some popular apps and demonstrate the feasibility of DroidLink.
\end{abstract}

\section{Introduction}
\input{Introduction}
\section{Background}
\input{Background}
\section{Approach}
\input{Approach}
\section{Implementation}
\input{Implementation}
\section{Evaluation}
\input{Evaluation}
\section{Discussion}
\input{Discussion}
\section{Related Work}
\input{Related}
\section{Conclusion}
\input{Conclusion}

\bibliographystyle{abbrv}
\bibliography{sigproc,inapp}

\end{document}

%% file: Introduction.tex
With the prevelance of mobile devices, apps have taken place the Web as the major entrance to access the Internet. The Web data is usually published and accessed via Web page that has a global URL and can be directly accessed from the URL. As a result, data can be crawled by search engines, and data from different domains can be linked via hyperlinks. In contrast, currently, data provided by apps is usually enclosed deeply inside apps so that they cannot be easily accessed by external users and other apps. To access the data of an app, users have to launch the app, locate the data by a series of actions such as searching and tapping, and then open the target ``content page'' containing the data. More severely, one app cannot access the data provided by another app unless the latter explicitly defines the interfaces and specifications. For example, opening Google Map app to locate the address of a restaurant in Yelp app requires that Google Map specifies the way how Yelp can access its internal data and status. However, such contract is usually made in quite ad-hoc and even case-by-case way. Data from different apps can hardly be linked together. As a result, it is argued that the apps become the ``isolated information islands'' and thus limit the ``\emph{mutual-composition}" between apps as well as the exploration of new features, functionalities, and even revenues~\cite{Li:IMC2015}.

The reason for preceding limitations can be various. At the early age of mobile apps, developers focus on only the features and functionalities of their own apps without having strong motivations to open and exchange data with one another via apps. In addition, under the context of mobile apps, there are no enforced standards such as HTML and URLs where the data inside apps can be organized and published to external. However, developers have been aware the importance of ``linked data'' that can produce more business opportunities. To realize the data interoperation, the concept of ``\textbf{deep link}'' is proposed to enable directly opening a specific page \footnote{If not specified, the term ``page'' refers to the content page of an app that contains the desired data. We also use ``page'' and ``content page'' interchangeably in this paper.} of an app from outside of this app with a uniform resource identifier (URI)~\cite{deeplink}. Many major Internet service providers such as Google, Apple, Facebook, and Microsoft are strong advocators of applying deep link. On the one hand, unlike the traditional Web pages, the apps are essentially ``black-box'' whose content and data cannot be accessed and indexed by external programs. App indexing and in-app search~\cite{GoogleAppIndexing,BingAppLinking,BaiduAppLink} become the focus of traditional Web search engines, such as Google, Bing, and Baidu, to support searching and requesting data inside apps. One basic requirement of app indexing is to enable opening the target app's content page that contains the matched data. On the other hand, some ``\textit{super}'' apps, such as Facebook~\cite{FacebookAppLinks} and WeChat~\cite{WeChat}, are desirable to become the entrance of mobile users by enabling transfers from themselves to other apps. Directly opening the target page of apps is necessary to provide better user experience. Therefore, deep link not only can bring convenience to mobile users, but is also beneficial to developers by increasing app's usage from search engines or super apps.

Technically, opening the deep link of a page inside an app can be supported by inter-process communication (IPC) of mainstream mobile OSes including Android, iOS, and Windows Mobile. However, as the first step, implementing the deep link for mobile apps and making the deep link fit for its usage scenario have to address two major challenges.

First, implementing deep links for an app may require a lot of manual refactoring efforts. A page of an app can have many dependencies that have to be initialized before being opened. For example, a restaurant-search app may construct an internal object of a restaurant in a restaurant page of showing detailed information. When transferring to a user-review comment page from the restaurant-info page, the app may use the constructed objects to request corresponding comments. If the app directly opens the comment page from outside without passing the restaurant-info page, the internal object will be missing. As a result, implementing the deep link for the comment page has to understand app logic, i.e., the dependencies among pages, to identify and remove all dependencies. As current mobile apps become quite complex with thousands lines of code, it is a non-trivial task at all.

Second, publishing deep links of an existing app may also require a lot of manual efforts. The deep link is useful when and only when it is published externally, so that the data inside an app can be accessible by third parties. Currently, there is no standard way of publishing deep links. Some solutions, such as Google App Indexing~\cite{GoogleAppIndexing} and Facebook App Links~\cite{FacebookAppLinks}, require that each page with a deep link must have a corresponding webpage. Such a webpage contains the descriptions of which data can be provided by the app. The deep link is published as a meta-data field in the webpage or is the same with the webpage's URL. For apps without corresponding webpages, other solutions, such as Wandoujia In-App Search~\cite{Wandoujia}\footnote{Wandoujia is a top Android app store in China, more information can be found in \cite{Li:IMC2015}}, require developers to submit the deep link and contents of the corresponding page to app stores. However, such a separate webpage or the contents have to be provided by the app developers, and inevitably increase the development and maintenance cost. In practice, very few developers are willing to adopt such a solution.

To address the preceding challenges, in this paper, we propose \emph{DroidLink} to automatically generating deep links for Android apps. We first design a deep link model that is suitable for automatic generation. Our model treats each instance of an activity (the basic component of Android apps) as a unit in which the deep link needs to associate with, abstracts executing a deep link as a series of activity transitions via Android Intent mechanism (the communication interfaces of activities), and publishes deep links as externally discoverable indexes.

Based on the model, DroidLink first analyzes the transition of activities of a given Android app and builds the Navigation Graph among pages inside an app with the help of dynamic and static analysis. Then for each activity that needs a deep link, in order to make the execution of deep links efficient, DroidLink computes a shortest transition path (which we denote as ``\textit{Shortcut}'') from the entrance of the app to the target activity, and maintains mappings between paths and the shortcuts. When packaging the app, a deep-link configuration file is automatically generated and deployed in the \texttt{APK} file, being able to accept deep link invocations from third-party apps or services.

In order to make deep links of an app discoverable, DroidLink keeps on crawling the app and visiting the deep link enabled activities. For each crawled instance, we compute the shortcut of the path to the page, and assigns an App URL to the instance. Then the content of the page is aggregated to generate an index of the page. Then the shortcut and the index are stored into a Deep Link Repository and published via a Web server. Third-party apps or services, such as Google and Facebook, can discover the deep links from the published Deep Link Repository. When a certain request to a deep link is issued, the target app traces back to the shortcut and triggers each activity transition sequentially with the corresponding Intent. Finally, the target page can be reached.

This paper makes the following contributions.
\begin{itemize}
    \item{We analyze the key requirements of deep links for mobile apps, and propose a Deep Link Model for Android apps that is suitable for automated generation.}
    \item{We propose DroidLink to automating the generation of deep links by creating a deep link enabled app and maintaining a index for deep links. The process requires minimal developer efforts and does not break the existed app structure.}
    \item{We evaluate the feasibility and efficiency of DroidLink on popular Android apps.}
\end{itemize}

The remainder of this paper is organized as follows. Section 2 describes the background of Android apps and deep link. Section 3 presents our approach to automating the generation of deep links. Section 4 shows the details of implementation. Section 5 evaluates our approach on popular apps. Section 6 discusses some issues and extensions of our approach. Section 7 highlights related work and Section 8 concludes the paper.

%% file: Background.tex
In this section, we first give some background knowledge of Android apps and deep link. Based on an example of how deep link is suggested to implement according to Google's App indexing, We analyze the challenges of implementing deep link for Android apps.

\subsection{Android Application}
\begin{figure}[t]
\centering
  \includegraphics[width=0.8\columnwidth]{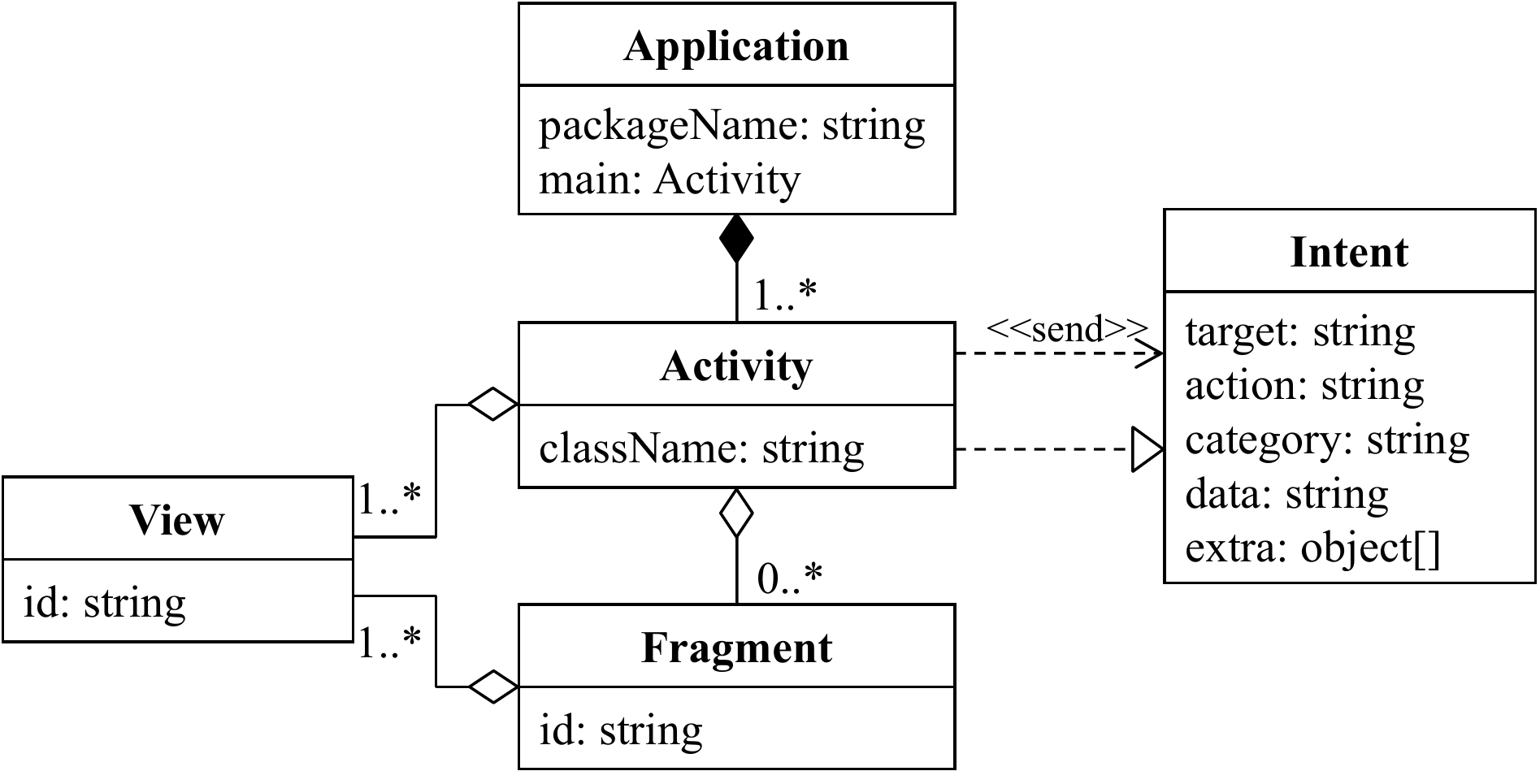}
  \caption{Android Application Model.}~\label{fig:activity}
\end{figure}

Deep link essentially provides a channel to access the internal state of an app. Figure~\ref{fig:activity} shows the structure of a typical Android app~\cite{AndroidGuide}. An app, distinguished by its \texttt{packageName}, usually consists of multiple Activities that are loosely bound to each other. An activity is a component that provides a screen with which users can interact in order to do something, such as dial the phone, watch a video, read a news, or view a map. Each activity has a unique \texttt{className} and is given a window in which to draw its user interface. One activity in an app is specified as the ``main'' activity, which is firstly presented to the user when the app is launched.

An activity has several \texttt{Views} to display its user interface, such as \texttt{TextView, ButtonView, ListView}, etc. To better organize views on screens with different size estate, Android provides the \texttt{Fragment} as a portion of user interface in an activity. Multiple fragments can be combined together to build a multi-pane UI. Fragments can also be reused in different activities. In a word, a fragment can be regarded as a modular section of an activity.

Activities are transited through \texttt{Intents}. An Intent is a messaging object that is used to request an action from another component, and thus essentially supports IPC communication at OS level. There are two types of intents: explicit intents that specify the target activity to start by the \texttt{className}; implicit intents that declare action, category and/or data which can be handled by another activity. Messages are encapsulated in the extra field of intent. When an activity $P$ sends out an intent $I$, the Android system finds the target activity $Q$ that can handle $I$, and then loads $Q$, achieving the transition from $P$ to $Q$.

\subsection{Deep Link}
The idea of deep link for mobile apps originates from the links in the Web. One significant difference between the Web and mobile apps is that each webpage has a global unique location (URL) and Web pages are accessible directly from anywhere else with the URL via a Web browser. Web users can input the URL of a webpage and click the Go button to open the Web page. They can either click through hyperlinks, transferring from one Web page to another.

However, app pages only have internal locations. Accessing an app page has to start from launching the app, transit through many pages, and finally reach the target page. For example, a user may find a favorite restaurant in a food app. Next time when the user wants to check the restaurant information, he/she has to launch the food app, search the restaurant again and then reach the page of the restaurant. In general, there is no way for the user to directly open the restaurant page even if he/she has visited before.

To solve the problem, deep link is proposed to enable directly opening a specific page of an app from outside of the app with a uniform resource identifier (URI). The greatest value of deep link is not limited in navigating users directly to a specific location of an app with a dedicated link, but supports other apps to be capable of accessing the internal state and data of an app and thus enables ``composition" of apps to explore more features, user experiences, and even revenues.

Since there is no standard definition or specification of deep link, we can compare the deep link with Web link and summarize three basic requirements that deep link should satisfy.
\begin{itemize}
    \item{\textbf{Executable.} With deep link, pages of an app can be directly opened without excessive user actions. This requirement is the fundamental motivation of deep link.}
    \item{\textbf{Identifiable.} Deep link should be expressed as a globally uniform identifier (like URL) that is attached to the deep linked pages. Pages of an app can be distinguished based on the deep link to them.}
    \item{\textbf{Discoverable.} Deep links of an app should be published in a way (like Web links through websites) to be discovered by third-party apps or services. Deep link is useful only when it is exposed to third parties.}
\end{itemize}

\subsection{An Example of Deep Link}
The business value of deep link has drawn a lot of attention, but lacks an enforcedly standard way. In practice, major Internet companies, such as Google, Facebook, Microsoft, and so on, usually design their own solutions. But all of these solutions share the similar concepts. Here we use the Google App Indexing~\cite{GoogleAppIndexing} as an example to illustrate how deep link is suggested to implement.

\begin{figure}[t]
\centering
  \includegraphics[width=0.5\columnwidth]{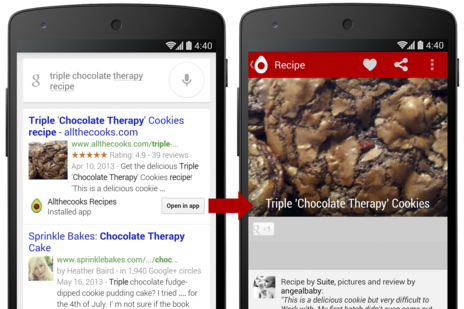}
  \caption{Google App Indexing.}~\label{fig:example}
\end{figure}
The goal of Google App Indexing is to put the apps in front of users who use Google Search. When a search result can be served from an app, users who have installed the app can go directly to the page containing the result. Figure~\ref{fig:example} shows an example of how App Indexing works for mobile users. When a user searches a term ``\textit{triple chocolate therapy recipe}'' on Google, one result comes from the `\textit{`Allthecooks Recipes}'' app. If the user clicks the button ``Open in app'', then the target app page of the search result is directly opened. App Indexing is actually a concrete usage scenario of deep link.

According to Google's developer guide, there are three steps to support such a kind of deep link for App Indexing.

The first step is to add some intent filters for activities showing the desirable pages. In the \texttt{AndroidManifest.xml}, intent filters declare the URL patterns that the app handles from inbound links. These URLs can be the same URLs used for the corresponding pages to the website. The following code snippet shows an intent filter for links to \url{http://www.examplepetstore.com} and \url{https://www.examplepetstore.com}:

\begin{scriptsize}
\begin{lstlisting}[language=XML]
<activity android:name="com.example.android.PetstoreActivity"
          android:label="@string/title_petstore">
   <intent-filter>
      <action android:name="android.intent.action.VIEW" />
      <category android:name="android.intent.category.DEFAULT" />
      <category android:name="android.intent.category.BROWSABLE" />
      <data android:scheme="http" />
      <data android:scheme="https" />
      <data android:host="www.examplepetstore.com" />
   </intent-filter>
</activity>
\end{lstlisting}
\end{scriptsize}

The second step is to add logic to handle intent filters. Once the system starts the app activity through an intent filter, it uses the data provided by the Intent to determine the app's view response. The \texttt{getDataString}() and \texttt{getAction}() methods are called to retrieve the data and action associated with the incoming intent. These methods can be called at any time during the lifecycle of an activity, but this operation should be generally done during early callbacks such as \texttt{onCreate}() or \texttt{onStart}(). The code below shows that the \texttt{onNewIntent}() method verifies the deep link format and displays the page content:

\begin{scriptsize}
\begin{lstlisting}[language=Java]
protected void onNewIntent(Intent intent) {
   String action = intent.getAction();
   String data = intent.getDataString();
   if (Intent.ACTION_VIEW.equals(action) && data != null) {
      String productId = data.substring(data.lastIndexOf("/") + 1);
      Uri contentUri = PetstoreContentProvider
                .CONTENT_URI.buildUpon()
                .appendPath(productId).build();
      showItem(contentUri);
   }
}
\end{lstlisting}
\end{scriptsize}

The last step is to declare a website associated to the app. Currently, the App Indexing requires the app to be indexed has a corresponding website. Google uses URLs discovered through Web indexing as the identifiers of deep links. Google search crawlers visit all the app pages whose URLs match the intent-filter patterns in the \texttt{AndroidManifest.xml} file.

From the preceding example, we can see that implementing deep link has to meet the three requirements have two significant challenges:
\begin{itemize}
    \item{\textbf{Manual refactoring to resolve the dependencies}. A deep link of an app page may have many dependencies that have to be initialized before opening the page. Developers have to manually refactor the way in which they implement the activities, and thus to make these activities  launched separately.}
    \item{\textbf{Manual maintenance to publish deep links}. The state-of-the-art solutions require deep-linked apps should have corresponding websites to publish deep links. However, many of these app developers do not have webpages. To support deep links, they have to manually create a new separate representation of the content to make deep links discoverable by third parties. Such task is not trivial and increases the development and maintenance cost.}
\end{itemize}

%% file: Approach.tex
\begin{figure*}[t]
\centering
  \includegraphics[width=1\columnwidth]{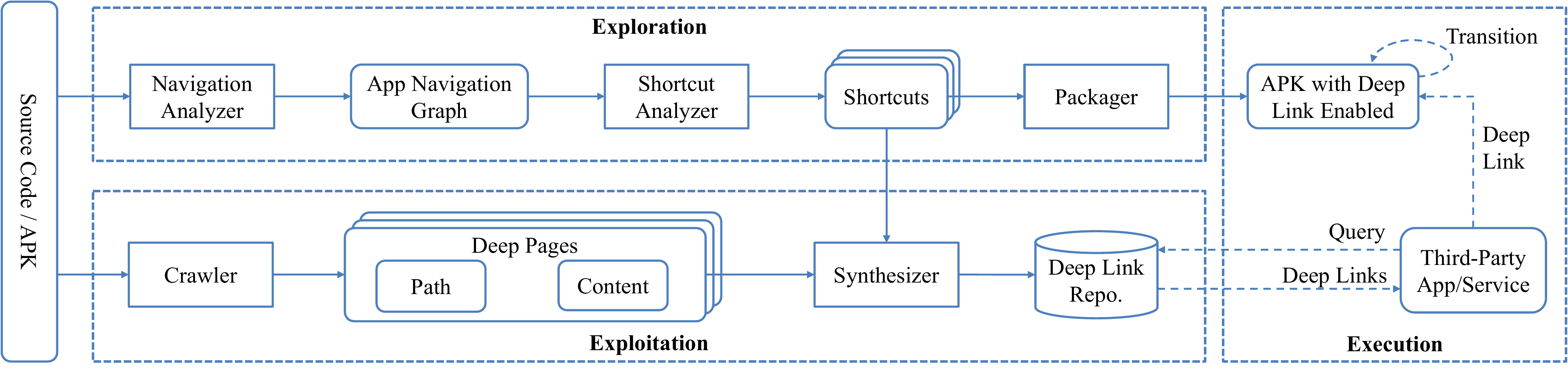}
  \caption{Overview of DroidLink.}~\label{fig:approach}
\end{figure*}
To address the two challenges of implementing deep link, we propose an automated approach DroidLink to generating deep links for Android apps. The generated deep link should be executable, identifiable, and discoverable and the generation should require minimal developer efforts and no additional modification to current Android OS. To this end, we design a deep link model that is suitable for automated generation. Based on the model, our approach consists of three processes (shown in Figure~\ref{fig:approach}): (1) \textbf{exploration} process that identifies how pages inside apps can be reached with least transitions; (2) \textbf{exploitation} process that publishes deep links of pages to make them discoverable to third parties; and (3) \textbf{execution} process that handles the actual query and invocation of deep links from third-party apps or services.

In the following, we present the details of our deep link model and each process, respectively.
\subsection{Deep Link Model}
To automate the process of generating deep links, we leverage the practices of the Web links and design a deep link model for Android apps. Our model regards the Android activity as a basic unit to be deep linked. Each instance of an activity is essentially a unique page, which contains information desirable for third parties.

To make the deep link executable, we abstract the execution of a deep link as a sequential path of activity transitions via intents, starting from the main activity of the app to the target activity pointed by the deep link.
\begin{Definition}[Activity Transition]
An activity transition $t(\mathcal{L})$ is triggered by an intent, where $\mathcal{L}$ is the combination of all the fields of the intent including action, category, data, and objects of extra.
\end{Definition}
Since an intent encapsulates several messages passed between two activities, we use a label set $\mathcal{L}$ to abstract an intent. Two intents are equivalent if and only if the label sets are the same.
\begin{Definition}[Path]
A path to an activity $\mathcal{P}_a$ is an ordered list of activity transitions $\{t_1, t_2, \dots, t_k\}$ starting from the main activity, where $k$ is the length of the path.
\end{Definition}
According to the definition, the activity transition $t_1$ is always the app launching intent that opens the main activity. The path $\mathcal{P}_a$ can ensure that all the internal dependencies are properly initialized before reaching the activity $a$.

To make the deep link identifiable, our model assigns an App URL to each deep-linked page. Pages of the same activity share a common schema.
\begin{Definition}[App URL]
An App URL $\mathcal{U}<host$, $target$, $hash>$ is an identifier for a instance of an activity, where the $host$ identifies an Android app, $target$ identifies an activity, and $hash$ identifies an instance.
\end{Definition}
The App URL is a combination among the identification of an app, an activity and an instance. Therefore, the App URL of a concrete page is globally unique.

To make the deep link discoverable, our model maintains a index of each deep linked page. The index is an abstraction of the content in the page so that it could be used by third-party apps or services to query desirable deep links to consume.

Finally, we present the formal definition of our deep link model based on the preceding notations.
\begin{Definition}[Deep Link Model]
A deep link is represented by a 3-tuple $<\mathcal{P}, \mathcal{U}, \mathcal{X}>$, where $\mathcal{P}=\{t_1, t_2, \dots, t_k\}$ is a path to the deep linked page, $\mathcal{U}$ is an App URL for the deep link, and $\mathcal{X}$ is an index of the deep linked page.
\end{Definition}
Our model ensures that the deep link is executable, identifiable, and discoverable with $\mathcal{P}$, $\mathcal{U}$ and $\mathcal{X}$, respectively.

\subsection{Exploration}
In the exploration process, given the source code or APK file of an app, the \textit{Navigation Analyzer} analyzes the activity transitions of the app and builds a\textit{ Navigation Graph} where each node represents an activity and edges represent transitions between two activities. Generally, there are multiple paths to reach an activity. Since our deep link model abstracts the execution of deep links as a path of activity transitions, the path should be as short as possible to reduce the execution time. Therefore, based on the \textit{Navigation Graph}, the \textit{Shortcut Analyzer} computes a shortest equivalent path, which we denote as a shortcut, for each path of an activity. At last, selecting the activities to be deep linked, the \textit{Packager} generates the \texttt{.APK} that supports invocation by deep link based on the shortcuts.

\subsubsection{Navigation Analyzer}
Activity transitions in an app could be very complicated. Since activities are loosely coupled that are communicated through Intent, there is no explicit control flows between activities. Therefore, we design a \textit{Navigation Graph} to abstract the activity transitions within an app.
\begin{Definition}[Navigation Graph]
A Navigation Graph $G$ is a directed graph with a start vertex. It is denoted by a 3-tuple, $G<V,E,r>$. $V$ is the set of vertices, representing all the activities of an app. $E$ is the set of directed edges. Each edge $e(v_1,v_2)$ represents an activity transition $t(\mathcal{L})$. $r$ is the start vertex.
\end{Definition}

In a Navigation Graph, the start vertex $r$ is the main activity of the app, which refers to the entrance of the app. The \textit{Navigation Graph} can have multi-edges, i.e., $\exists e_1,e_2\in E, v_{start}(e_1)=v_{start}(e_2)~and~v_{end}(e_1)=v_{end}(e_2)$, indicating that there could be more than one transitions between two activities. We enforce that each node in $V$ should be reachable from the start vertex $r$. Such a definition means that we consider only the activities that can be reached from the main activity. In fact, some activities can only be reached by system events such as receiving a message. We do not take into account these system-triggered activities in this paper. In addition, it should be noted that the \textit{Navigation Graph} can be cyclic.

Algorithm~\ref{algo:ang} describes how \textit{Navigation Graph} is constructed. The algorithm follows a dynamic analysis manner to actually exercise the app to find activity transitions. We use a parameter $depth$ to indicate the longest activity transitions to examine. Initially, the main activity is set to the start node $r$ (Line 1) and added to $V$ (Line 2). We maintain a list $ActSet$ for each depth to store the activities that have been explored already (Line 3). For each exploration depth, we iterate each activity that has been discovered (Line 5). Then we exercise the activity to traverse all transitions to other activities (Line 6). Currently, we consider only the activities that can be reached by click and scroll actions. For each of the reached activity, if it does not exist in current graph, then we add it to $V$ and to the exploration list of next depth (Line 8-11). We do not recursively examine the activities that have already been added in the graph. The algorithm finishes when all the depths have been explored.
\begin{algorithm}
    \caption{Construction of Navigation Graph}
    \label{algo:ang}
    \SetAlgoLined
    \KwIn{$app$ to be deep linked, exploration $depth$}
    \KwOut{Navigation Graph $G<V,E,r>$}
        $r\leftarrow app.main\_activity$;\\
        $V$.add($r$);\\
        $ActSet(0)$.add($r$);\\
        \For{$d\leftarrow~0$ to $depth$} {
            \ForEach{$ori\in ActSet(d)$} {
                $TarAct\leftarrow$getTransitions($ori$);\\
                \ForEach{$tar\in TarAct$} {
                    \If{$tar\notin V$} {
                        $V$.add($tar$);\\
                        $ActSet(d+1)$.add($tar$);
                    }
                    $E$.add($ori$, $tar$, getIntent($tar$));
                }
            }
        }
\end{algorithm}

\subsubsection{Shortcut Analyzer}
\begin{figure}[t]
\centering
  \includegraphics[width=0.7\columnwidth]{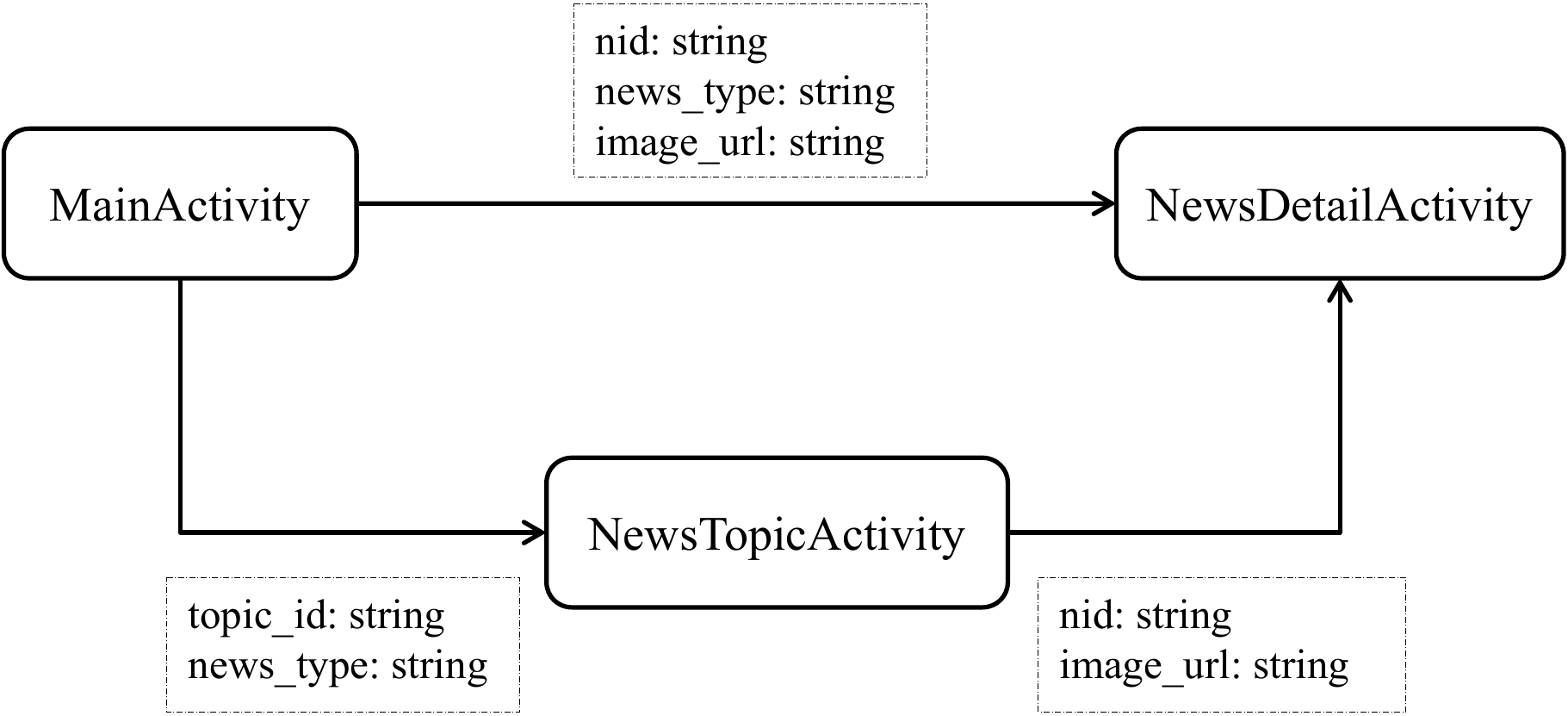}
  \caption{Example of Wallstreet News App.}~\label{fig:wallstreet}
\end{figure}

An activity can have several paths to reach. Figure~\ref{fig:wallstreet} shows the activity transitions in ``\textit{Wallstreet News}'' app. If we want to reach a news page of \texttt{NewsDetailActivity}, there are two paths. One is to navigate directly from the \texttt{MainActivity}. The other is to switch to the topic page of \texttt{NewsTopicActivity} and then navigate to the news page. Obviously, the former path is shorter than the latter one. Since our approach uses the activity transitions to execute the deep link, the path should be as short as possible to reduce the execution time. However in some circumstances, a shorter path to an activity cannot cover every instance. For example, medium pages on the path may depend on some internal variables, thus they can only be reached via a longer path in which these variables are assigned.

We define that path $p_1$ can replace path $p_2$ if and only if $\mathcal{L}_{p_1}\subset \mathcal{L}_{p_2}$. Here $\mathcal{L}_{p_j}$ is the combination of all the labels in the path $p_j$: $\mathcal{L}_{p_j}=\cup\{\mathcal{L}(t_i)|t_i\in p_j\}$. For example, in ``\textit{Wallstreet News}'' app, path 1 from \texttt{MainActivity} to \texttt{NewsDetailActivity} needs three labels, ``\textit{nid}'', ``\textit{image\_url}'', and ``\textit{news\_type}''. We can see that all these three labels are contained in path 2 which passes the \texttt{NewsTopicActivity}. So we can use path 1 to replace path 2.

With the definition of path replacement, for a given path we define the possible shortest path as a Shortcut. We should use the shortcut to replace the path to an activity for shorter execution time.
\begin{Definition}[Shortcut]
A Shortcut of a path $\mathcal{C}(p)$ is the shortest path that can replace path $p$.
\end{Definition}

Algorithm~\ref{algo:shortcut} is for finding shortcuts in Navigation Graph $G$. For every vertex in the graph, get its paths (Line 2) and sort the paths by their length in ascending order (Line 3). Then enumerate every path in the path list, to find the shortest path that can replace it (Line 4-12), if they fulfill the requirement of label containing (Line 7). Due to the increasing sequence, the first available path we get is the shortest one. We store it in a two-dimension map structure (Line 8).
\begin{algorithm}
    \caption{Shortcut computation.}
    \label{algo:shortcut}
    \SetAlgoLined
    \KwIn{Navigation Graph $G<V,E,r>$ }
    \KwOut{Shortcut $\mathcal{C}$}
        \ForEach{$v\in V$} {
            $Plist\leftarrow ANG.path(v)$;\\
            sort\_by\_length($Plist$);\\
            \For{$i\leftarrow 1~to~Plist.length$} {
                $Shorcut[<v, p_i>]\leftarrow p_i$;\\
                \For{$j\leftarrow 1~to~i-1$} {
                    \If{$\mathcal{L}_{p_j}\subset\mathcal{L}_{p_i}$} {
                        $Shorcut[<v, p_i>]\leftarrow p_j$;\\
                        break;
                    }
                }
            }
        }
\end{algorithm}
\subsubsection{Packager}\label{sec:packager}
After computing the shortcuts, the next step is to create the target \texttt{APK} that supports deep link. Note that not all of the activities need to be deep linked in practice. Developers may want to share only some data of their apps. Therefore, the \textit{Packager} lists all the explored activities and allows developers to check which ones need deep link support.

Then the Packager generates an abstract schema of App URL for each selected activity. We employ the format of ``\textit{http://host/target?hashkey}'' as the representation of the App URL. We use the reverse string of the packageName for the host field and the full className of the activity for the target field. All the instances of an activity share the similar prefix before ``?''. With the schema of App URL, we can generate intent filters in AndroidManifest.xml file to handle the corresponding deep link.
\begin{figure}[t]
\centering
  \includegraphics[width=0.6\columnwidth]{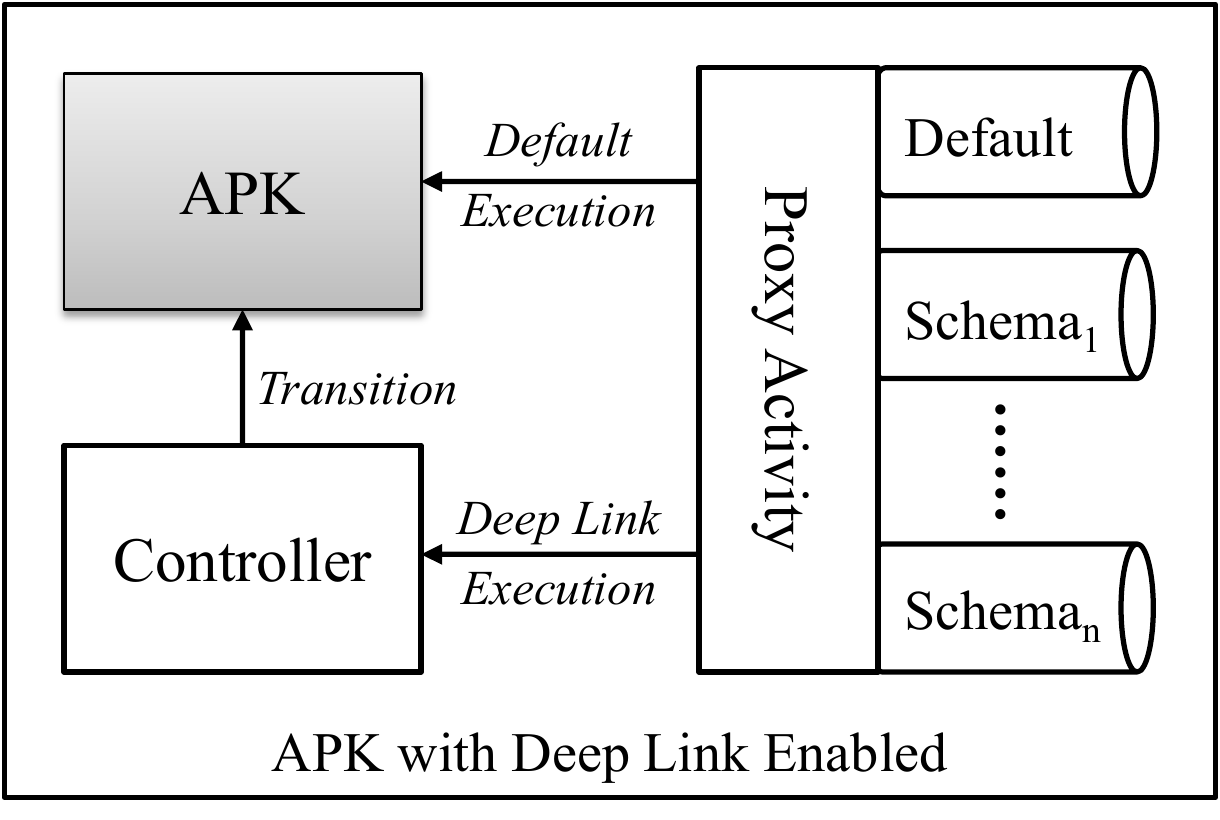}
  \caption{Structure of the APK with deep link enabled.}~\label{fig:package}
\end{figure}

Finally, an \texttt{APK} with deep link supported is created. Figure~\ref{fig:package} depicts the structure of the created \texttt{APK}. We leverage a proxy architecture to realize minimal refactorings to the original app. A \textit{Proxy Activity} is used to handle all the incoming requests. The Proxy Activity is configured to intent filters that conform to the URL schemas. When an intent is passed to the \textit{Proxy Activity}, if the intent matches one of the schemas, the \textit{Proxy Activity} informs the Controller to execute the deep link. The \textit{Controller} communicates with the original \texttt{APK} and instructs the app to transit through activities according to the path. If the incoming intent cannot match to any of the schemas, it is then forwarded directly to the original \texttt{APK} for default execution.

\subsection{Exploitation}
The exploration is an automatic process that involves two major algorithms, the Crawler and the Synthesizer. The Crawler keeps on visiting the instances of activities that support deep link. For each instance, which we denote as a deep page, the Crawler outputs a concrete path from the main activity as well as the content of the page. Given all the paths and contents extracted by the Crawler, the Synthesizer uses the shortcut to compute a shortest sequence of each concrete path and generates the corresponding App URL. In order to reveal the content of the deep pages and make deep links discoverable, the Synthesizer draws up an abstract for contents of each path as a index. The deep links, including their App URL, concrete paths and descriptors, are stored into a Deep Link Repository that can be accessed by users.

\subsubsection{Crawler}
The Crawler takes three steps to process each single deep page.

1. Acquire a path from the Deep Link Repository.

2. Start from home page and execute every transitions in the path, which brings the Crawler to the initial state of the target deep page.

3. Fully exploit the deep page to discover new paths and contents.

Occasionally, the deep page is simple and the Crawler can directly discover all the paths and contents in the initial state. However, there are also more complicated deep pages where more contents and paths are hidden until implementing some designated user operations. As a result, the Crawler should exploit more states of each deep page by detecting and executing every possible operations. To achieve such objective, the Crawler uses Page Sate Transition Graph to record and exploit each state of every deep page.

\begin{Definition}[Page State Transition Graph]
A Page State Transition Graph is denoted by a 7-tuple, $PSTG(V,E,s,\Omega,C,\xi,\zeta)$, where V is the set of vertices, E is the set of (direct) edges, s is the start vertex, $\Omega$ is the set of user operations, C is the set of contents,
$\xi:E\rightarrow\Omega$ is a function assigning a operation to each edge and
$\zeta:V\rightarrow C$ is a bijection mapping every vertices to their contents.
\end{Definition}

The definition shows that we identify each state of a deep page by its content. If an operation doesn't result in any change of the content, there won't be any new state inserted into the Graph.

The Crawler dynamically generates the PSTG from the original state initiated by the path. Then, it repeatedly explores each state in the current PSTG, discovers new paths or states, and updates the PSTG and notifies the Synthesizer.

\begin{algorithm}
    \caption{Crawler}
    \label{algorithm crawler}
    \SetAlgoLined
    \KwIn{An deep page in its original state}
    \KwOut{The contents and links in the deep page}
        INITIAL $PSTG$ with the original state\\
        \ForEach{$state_i \in PSTG$}
        {
            \ForEach{$operation_j \in state_i$}
            {
                execute $operation_j$ \\
                \If{discover a new deep page}
                {
                    notify the Synthesizer with the path
                }
                \Else {
                    \If{$currentstate \neq state_i$ \\
                    and $currentstate \notin PSTG$}
                    {
                        update $PSTG$ with the $current state$ \\
                        ~~the new $content$ and the $operation_j$
                    }
                }
                recover $state_i$
            }
        }
\end{algorithm}

The Crawler is focused on exploiting one state at a time. The $state_i$ in line 2 is the state being processed. In line 3 and 4, it selects and invokes an executable operation from $state_i$. After every execution the Crawler tests if a new deep page occurs. If so, it has detected a new path that links from home Page to the current page, which should be sent to the Synthesizer in line 6. If not, the Crawler then inspects the current contents in the deep page. Any change in the contents indicates a newly discovered states. So the Crawler should update the current PSTG in line 11 and 12. After the executions above, the Crawler should recover the $state_i$ in line 15, to maintain the original state for the next operation.

\subsubsection{Synthesizer}
The Synthesizer collects the paths and the contents from the Crawler to generate App URLs and descriptors with the information provided by the exploration process.

The paths generated by the Crawler contains the steps from the home page to the target page, but there could be redundant steps that can be eliminated without interfering the functioning of the deep links. The Shortcut mapping function defined in the Exploration Process can help the Synthesizer to shorten the paths and generates more efficient deep links. To improve the utility of App URLs, the Synthesizer extracts a brief abstract from the contents of each deep page as a descriptor, which can provide a clue for common users who may not be very familiar with the specific Application and the deep link system. The major steps of the Synthesizer are described below.

1. Serialize the path from the home page to the destination page.

2. Generate hash code from the serialized path.

3. Attach the hash code to the package and the activity name to generate App URL

4. Extract an abstract as an index of the URL from the contents in the PSTG of the destination page, according to the size, color, position or other attributes of the texts.

5. Register the path, App URL and the index to the Deep Link Repository.

\subsection{Execution}
In the execution process, third-party apps or services, such as Google App Indexing, are able to discover desirable deep links of an app from the published \textit{Deep Link Repository}. The repository is deployed on a Web server and each deep link is in the form of a common Web URL. When an HTTP request is sent to the URL, the corresponding descriptor is returned. There are two typical usage scenarios of deep link discovery. On the one hand, app developers can provide a directory of the \textit{Deep Link Repository} for search engines, such as Google, Bing, and Baidu. So crawlers of search engines are able to retrieve deep links and the corresponding contents in order to build app index. On the other hand, those super apps, such as Facebook, provide interfaces for other apps to deeply integrate. App developers can submit deep links to these interfaces in order to promote the usage of their apps.

When a user actually clicks a deep link from third-party apps or services, e.g. the user would like to see details of a search result, the request to a deep link is emitted to the Android system and the target app can receive the request. As shown in Section~\ref{sec:packager}, the \textit{Proxy Activity} is launched and indicates the \textit{Controller} to handle the deep link execution. The \textit{Controller} contacts with the server that publishes the deep links to obtain the corresponding concrete paths of the\textit{ App URL}. Then the Controller issues the activity transitions in the path one by one to the original \texttt{APK}. Finally, the deep page identified by the \textit{App URL} is reached.

%% file: Implementation.tex
In this section, we present the implementation details of our approach. Our current implementation targets the Android platform. We use the instrumentation test framework provided by Android SDK to implement the automating process. Android instrumentation can load both a test package and the app under test (AUT) into the same process. Therefore using this framework, we are able to inspect and change the running environment of an app, such as retrieving view components of an activity, examining the intent that triggers an activity transition, as well as triggering a user action on the target app.

The exploration process is implemented as a tool of dynamic analysis. The tool takes the source code or \texttt{APK} file as the input, automatically generates an instrumentation test project that traverses the app, records the activity name and intent labels, and dynamically constructs the \textit{Navigation Graph}. We restrict the traverse depth for efficiency purpose. When the construction of \textit{Navigation Graph} completes, the shortcut analyzer works to compute shortcuts for all the paths of all nodes in the graph.

In the exploitation process, we use a virtual machine running Android x86, serving as crawlers. We set up a database to store deep links and paths. The crawler is also implemented by the instrumentation test framework. The crawler gets a path of a page to be exploited from the database, then calls a series of \texttt{startActivity()} command to transit to the target page, obtain all view objects by \texttt{getCurrentViews()} and traverses all the views using \texttt{clickonview()}. If the crawler find a new page, it records the serialized intent and computes a hash to generate the App URL, then stores the App URL and the descriptor into the database. We implement the page descriptor in the format of HTML and they can be deployed on any Web server.

To generate the \texttt{APK} with deep link enabled, the \textit{Controller} is actually implemented as a test case of the instrumentation test to execute the traversing methods. We add a proxy activity that receives external requests of deep links and a \texttt{AndroidManifest.xml} file is also generated. When the proxy activity receives a deep link request, it match the \textit{App URL} to the schema filter and map it back into a series of activity transition data. Then the \textit{Controller} launches the instrumentation test to get to the target page.

%% file: Evaluation.tex
In this section, we evaluate the feasibility and efficiency of our approach. Our evaluation addresses the following research questions.
\begin{itemize}
\item {\textbf{RQ1}: Is our approach feasible to automatically generate deep links for Android apps, even without source code?}
\item {\textbf{RQ2}: How efficient is our exploration process to cover as many activities as possible?}

\item {\textbf{RQ3}: How efficient is our exploitation process to publish deep linked pages?}
\end{itemize}

\subsection{Experiment Setup}
\begin{table}[t]\scriptsize
\centering
\caption{Selected apps for evaluation.}\label{table:apps}
\begin{tabular}[t]{llll}
 \hline
  \textbf{App} & \textbf{Version} & \textbf{Category} & \textbf{Description} \\
  \hline
  Reddit & 1.0.3 & News \& Magazines & Get the latest news and trends on the Internet first. \\
  %\hline
  Yahoo & 6.6.10 & News \& Magazines & Track developing stories \& stay informed with millions of people. \\
  %\hline
  eBay & 4.2.1.1 & Shopping & An e-commerce app. \\
  %\hline
  AliExpress & 4.8.6 & Shopping & An online shopping marketplace. \\
  %\hline
  Yelp & 8.8.0 & Travel \& Local & A local guide to finding just the place to eat, shop, and play. \\
  \hline
\end{tabular}
\end{table}
We select five popular apps on Google Play for our evaluation (shown in Table~\ref{table:apps}). There are two news apps, two shopping apps, and one travel app. All these apps belong to information-intensive categories so that adding deep links to their content pages is of great value to bring more traffic and conversions to the apps.

Our evaluation is conducted on a virtual machine with 2 CPU cores and 2G memory, running Ubuntu 12.04 OS. For dynamic analysis of Android apps, we set up an emulator of Android 4.3 with 1 CPU core and 1G memory to run the Android apps. We deploy our exploration tool and exploitation systems on the virtual machine.

\subsection{Case Study}
To study \textbf{RQ1}, we select the ``Reddit'' app to demonstrate the feasibility of our approach.
\begin{figure}
\centering
  \includegraphics[width=1\columnwidth]{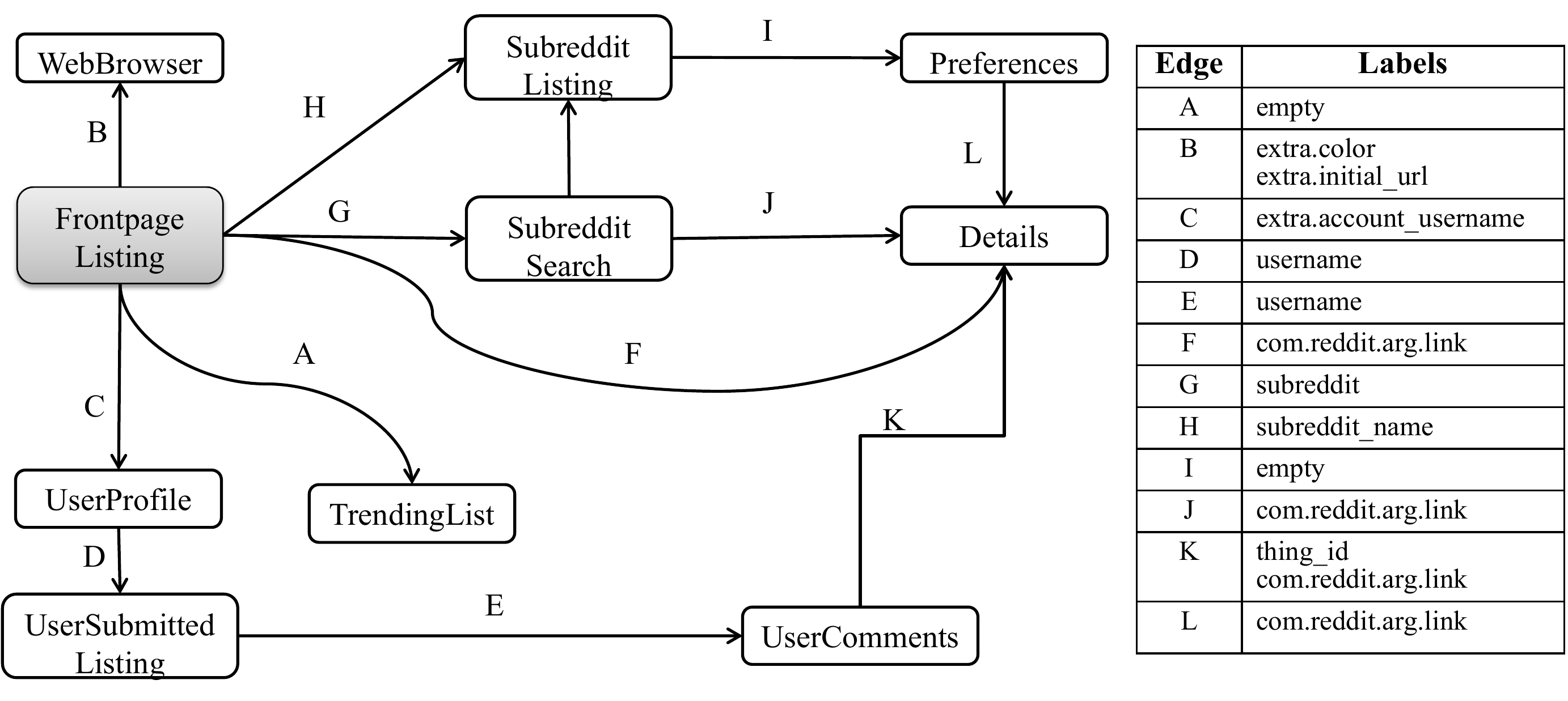}
  \caption{Nagivation Graph of Reddit.}~\label{fig:demograph}
\end{figure}
First we explore the internal structure of the app and generate the Navigation Graph with the depth of four. Figure~\ref{fig:demograph} shows a partial graph, in which each node represents an activity and each directed edge represents an activity transition. There are 10 activities in the graph and the main activity of the app is ``FrontpageListingActivity''.

\begin{table*}[t]
  \centering
  \caption{Paths to the DetailsActivity.}\label{table:pathexample}
  \begin{footnotesize}
\begin{tabular}{l|c|l}
 \hline
  Path & Length & Labels \\
  \hline
  FrontpageListing$\rightarrow$Detail & 1 & (1)arg.link \\
  FrontpageListing$\rightarrow$SubredditListing$\rightarrow$Detail & 2 & (1)subreddit\_name; (2)arg.link \\
  FrontpageListing$\rightarrow$SubredditListing$\rightarrow$PreferencesActivity$\rightarrow$Detail & 3 & (1)subreddit\_name; (2)arg.link \\
  \tabincell{l}{FrontpageListing$\rightarrow$UserProfile$\rightarrow$UserSubmittedListing$\rightarrow$\\~~~~~~UserComments$\rightarrow$Detail} & 4 & (1)account\_username; (2)username; (3)arg.link \\
  \hline
\end{tabular}
\end{footnotesize}
\end{table*}

Next, we use the graph to generate the shorcuts. Take the ``DetailsActivity'' as an example. In the graph, there are totaly four paths to reach the DetailsActivity (as shown in Table~\ref{table:pathexample}). The shortest path is to directly jump from the main activity. Through anaysis, we can find that all these four paths can provide the labels that the shortest path requires, which means the shortest path is the shortcut for all the four paths.

Then, we exploit the content of Reddit using the Crawler, and generate the concrete paths of every activity instance. Afterwards the Synthesizer uses the shortcut to gernerate a shortest sequence of activity transitions, hash the shortest sequence and generate a deep link by the App URL schema. For example, an instance of the path towards detail page includes:

\begin{scriptsize}
\begin{lstlisting}
FrontpageListingActivity
  Intent {cmp=com.reddit.frontpage/.FrontpageListingActivity}
SearchActivity
  Intent {cmp=com.reddit.frontpage/.SearchActivity}
DetailActivity
  Intent {cmp=com.reddit.frontpage/.DetailActivity (has extras)}
  com.reddit.arg.link=com.reddit.requests.models.v1.Link@529a13f8
\end{lstlisting}
\end{scriptsize}

We match the path with the shorcut of it, and get a shortest sequence:

\begin{scriptsize}
\begin{lstlisting}
Intent {cmp=com.reddit.frontpage/.DetailActivity (has extras)}
com.reddit.arg.link=com.reddit.requests.models.v1.Link@529a13f8
\end{lstlisting}
\end{scriptsize}

\begin{figure}
\centering
  \includegraphics[width=0.5\columnwidth]{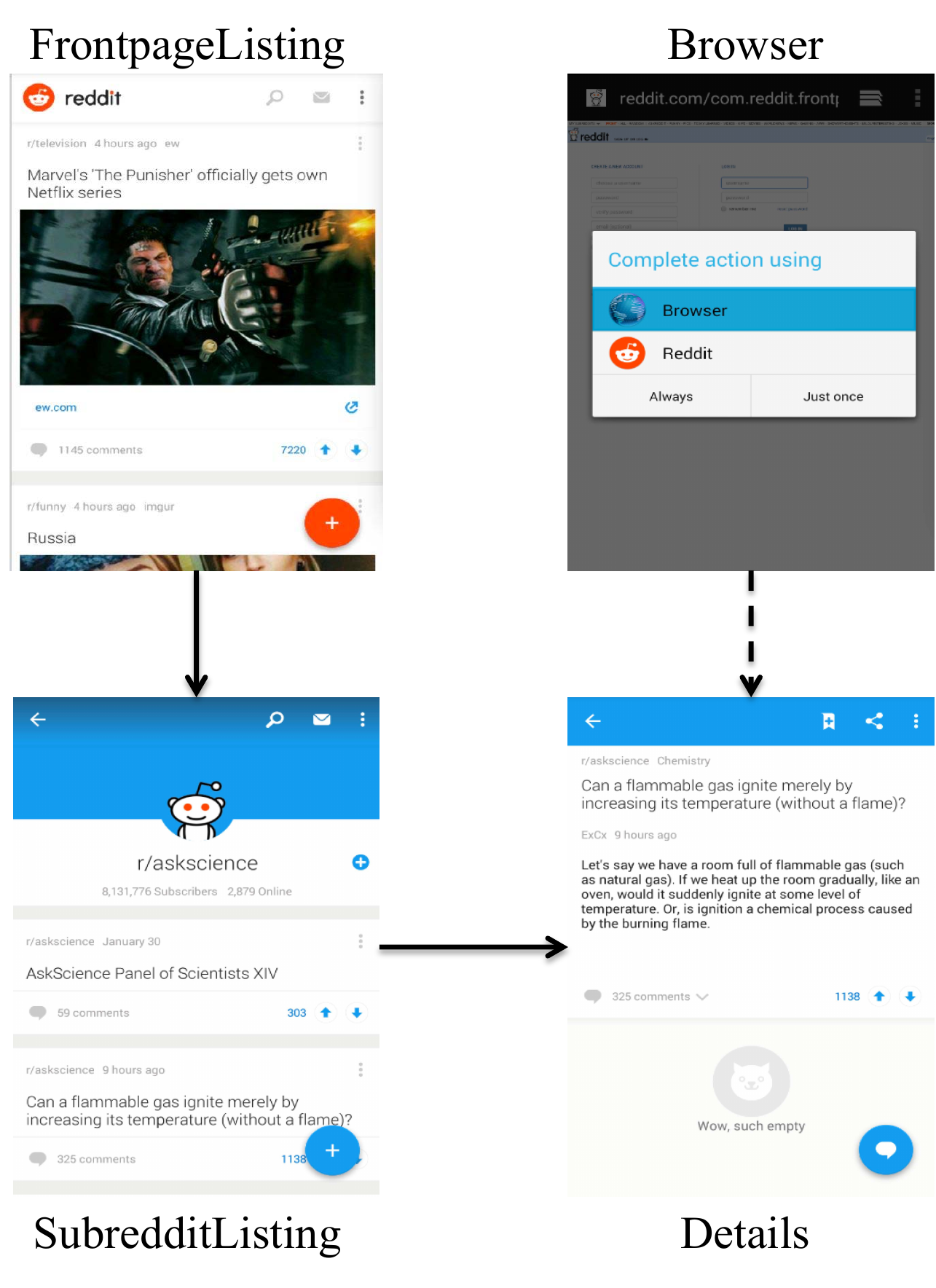}
  \caption{Execution of a deep link for Reddit.}~\label{fig:demo}
\end{figure}

Finally, we extract the content of this page and store in the Deep Link Repository. Combining the page with the URL, we publish it on our server. The App URL is \begin{scriptsize}$$http://reddit.com/com.reddit.frontpage.DetailActivity?5515207922751745125$$\end{scriptsize} We use the generated the Reddit APK with Deep Link enabled to test our generated deep link. The result is showed in Figure~\ref{fig:demo}. By visiting the App URL we published before on a mobile browser, the Reddit app can directly open the target page. The original operations need three steps (the solid line), while the deep link offers a direct jump to the target page (the dotted line).

\subsection{Evaluation of the Exploration}
The goal of the exploration process is to discover more possible paths between activities to generate more efficient shortcut mappings. Obviously, the sufficiency of paths detected can be influenced by the max depth of our exploration process. Normally, deeper exploration brings more comprehensive Navigation Graph. However, the complexity of computation also grows when the depth increases. As a result, identifying the appropriate exploration depths is a important task as raised by RQ2.

\begin{figure*}[t]
  \centering
  \subfigure[Number of explored activities.]{
    \label{fig:exploration:a} %% label for first subfigure
    \includegraphics[width=0.45\textwidth]{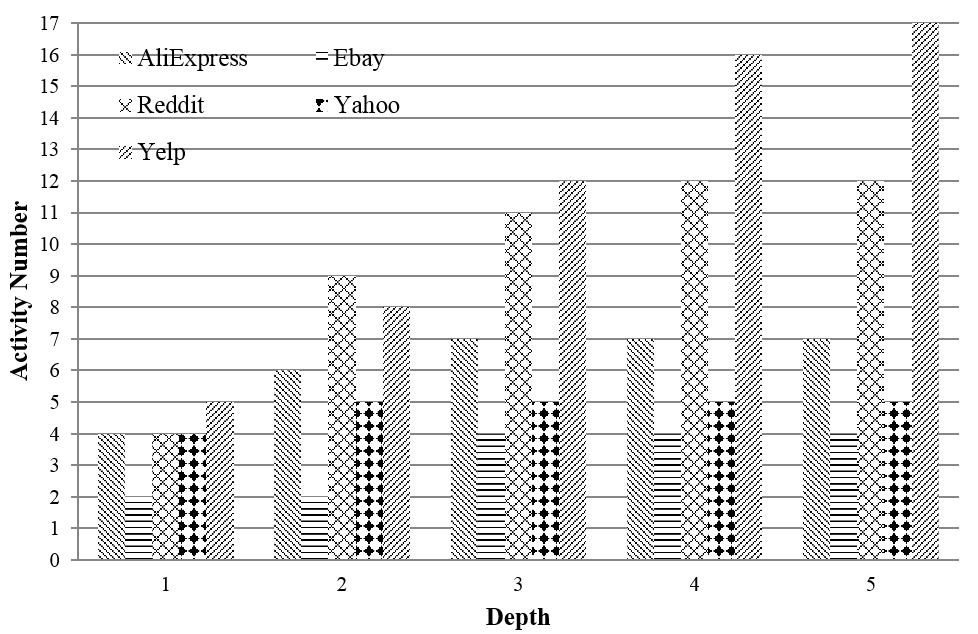}}
  \subfigure[Computation time of the exploration process.]{
    \label{fig:exploration:b} %% label for second subfigure
    \includegraphics[width=0.45\textwidth]{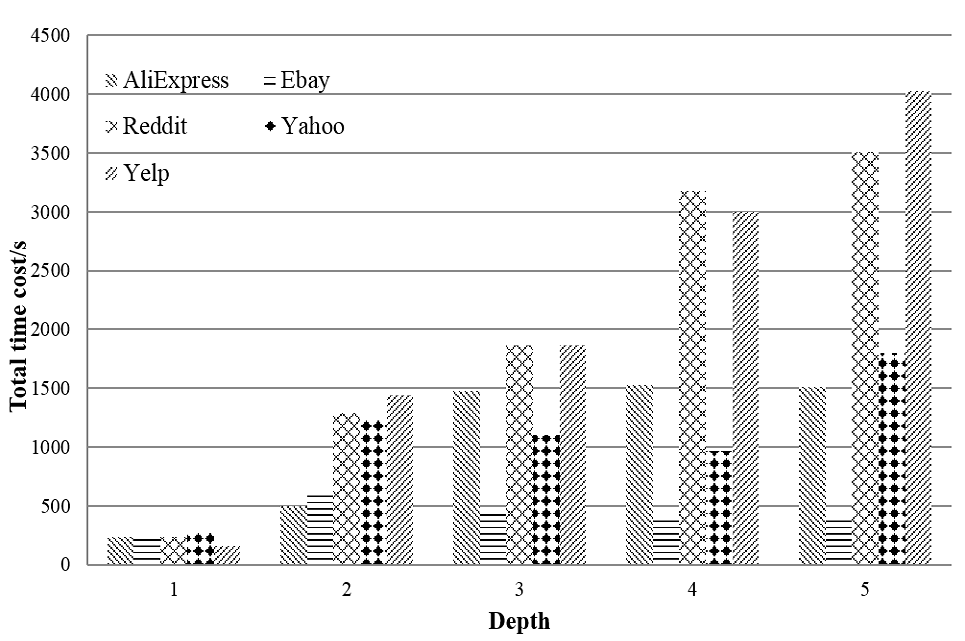}}
  \caption{Results of the exploration evaluation.}
  \label{figure:exploration} %% label for entire figure
\end{figure*}

To evaluate the efficiency of exploration, for each of the 5 selected apps, we execute 5 rounds of navigation analysis, in which the max depths varies from 1 to 5, respectively. We use the number of detected activities to indicate the completeness of the generated Navigation Graph. To measure the complexity of computation, we record the time consumed by each round. The results are shown in Figure~\ref{figure:exploration}.

As shown in Figure~\ref{fig:exploration:a}, there is a difference among total amounts of discovered activities: AliExpress, Ebay and Yahoo have relatively smaller sizes of activity set, which is 7, 4 and 5 respectively, while Reddit and Yelp have larger ones, which is 12 and 17. This feature also correlates the characteristic of convergence. The former three applications converges more early at about the second or third round. On the contrary, Reddit converges at the fourth round and we don't have enough evidence to tell if Yelp has already converged at the last round.

The time that elapses on each round of different apps also follows the same pattern, as shown in Figure~\ref{fig:exploration:b}. The time consumed by Ebay and Yahoo stops growing in the second round, and AliExpress stops at the third round. The time consumed by Reddit seems to slow down the growing in the last two rounds, but the time of Yelp still increases steadily until the evaluation ends. There is also an abnormal re-elevation on the time consumed by Yahoo in the Fouth round. But apparently, there isn't any growth in the number of new detected activity. These observations indicate that the time and amount of detected activity in navigation normally converges simultaneously when the depth grows, but larger depth may lead to more time consumed without gains in the completeness of the Navigation Graph. It is proper to choose smaller depth for relatively simple apps, such as AliExpress, Ebay and Yahoo. For more complex apps such as Reddit and Yelp, the depth should be also larger. For AliExpress, Ebay and Yahoo, the ideal depth should be 2 or 3, while for Reddit 4 is sufficient. Deficiency in depth may generates incomplete Navigation Graph and Shortcut mappings that can cause the paths in the Deep Links to have redundant steps. But surplus in depth may cause increase in time that is solely noneffective.

\subsection{Evaluation of the Exploitation}
To study \textbf{RQ3}, we continue to crawl the five selected apps for five hours. Afterwards, we first analyze the number of pages our system exploits. Then we study the distribution of activities from which these pages comes. Results are shown in~\ref{figure:exploitation}.

Figure~\ref{fig:exploitation:a} shows the relationship between the number of exploited deep links and the consumed time. As illustrated, the line has a linear-shape, indicating the Crawler keeps a rather steady pace to exploit deep linked pages. Since the Crawler adds only new pages to the repository, so the generation of the deep linked page reflects the update frequency of the app. In addition, we can observe that for different apps, the time used for a new deep linked page is different. For example, the average time for Reddit to publish a page is 37.6s, while Yelp's is 84.1s. The difference is mainly caused by the depth of the page. Most of Reddit's pages can be crawled from the front page, thus dramatically decreasing the time cost. Besides, an app with a high frequency of update can have a low time cost for deep link generation.

Next we analyze the distribution of activities to which these deep linked pages belong. Figure~\ref{fig:exploitation:b} shows the result. We sort the activity by the number of pages that are the instance of the activity, and display the proportion of each activity on the x-axis, while the page proportion is displayed on the y-axis. As we can observe in the figure, in all these apps, most pages come from only a small set of activities. For example, in Aliexpress, over fifty percent of exploited pages are the instances of ProductDetailActivity, which is the terminal of content display.

%% file: Discussion.tex
In this section, we discuss the limitations of our approach, as well as applications and implications based on our work.

Our approach currently replies on dynamic analysis of Android apps to build the \textit{Navigation Graph}. However, there are some activities that our \textit{Navigation Analyzer} cannot reach. As a result, the explored navigation graph is not completed. For example, some activities are triggered only in user-specific scenarios, e.g., require user login to show user-related data. These activities can be reached only with user identities. Currently, we do not consider these activities and our approach only takes in public information pages. One possible solution is to synthesize static analysis to compute a intermediate structure of the app. 

Our approach uses Intent to distinguish activity transitions. Different instances of Intent could transit to different instances of an activity. We also assume that activity transitions can be re-executed by re-sending the Intent. However, in practice, some activity transitions do not instantiate different Intents. Instead, they use public objects to pass messages. In such a case, the Intent instances of activity transitions are always the same so that our approach cannot apply directly. One possible solution is to leverage static analysis to find the data related to activity transitions. This solution is like taint analysis to some extent.

The execution of deep link may have security and privacy issues. Currently, our approach allows the developers to decide which activities to support deep link. For simplicity, we assume that all the instances are permitted to be deep linked. Furthermore, we establish index to make the deep-linked data discoverable. As a result, there can be risks to leak sensitive data and be attacked by hackers. We plan to enhance security and privacy with some preliminary efforts~\cite{Xie:USENIX2013}.

Indeed, the idea of deep link is quite new but is showing its importance to break up the barriers of current ``isolated'' apps. Deep link is now developed in a very ad-hoc and tedious way. Our work in this paper makes the preliminary effort to accelerate the productivity of deep links in a more automated fashion. The model and techniques can establish the fundamental supports for Android app developers who are willing to open their data to others so as to provide more features and optimized user experiences, and even gain more users as well as revenues from the data exchange. Although the techniques in this paper is for Android apps, the basic principle itself can be customized and applied to other platforms such as iOS and Windows Mobile.

%% file: Related.tex
\begin{figure*}[t]
  \centering
  \subfigure[Time consumption of exploitation.]{
    \label{fig:exploitation:a} %% label for first subfigure
    \includegraphics[width=0.4\textwidth]{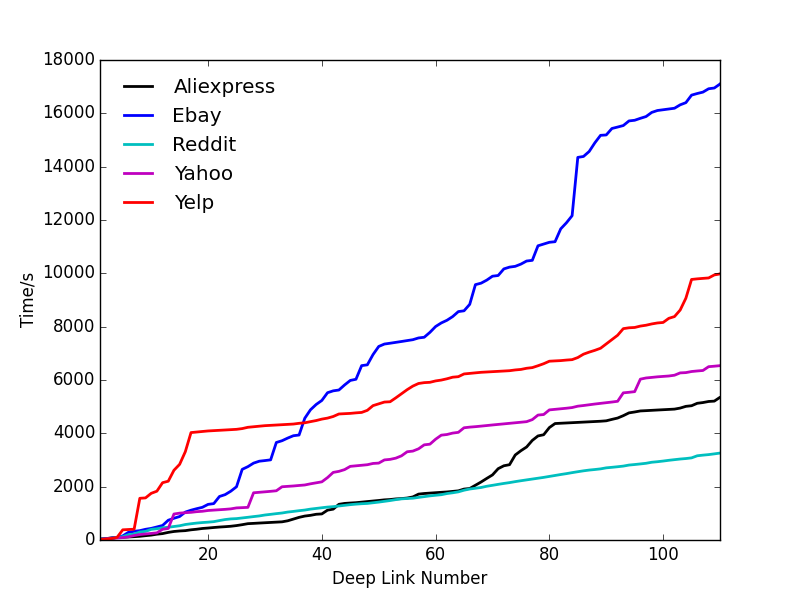}}
  \subfigure[Distribution of activities of exploited pages.]{
    \label{fig:exploitation:b} %% label for second subfigure
    \includegraphics[width=0.4\textwidth]{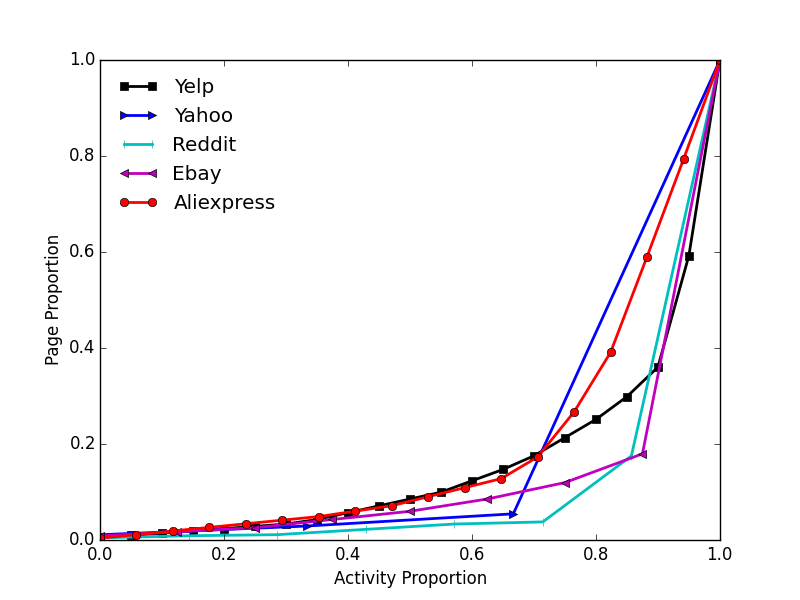}}
  \caption{Results of the exploitation evaluation.}
  \label{figure:exploitation} %% label for entire figure
\end{figure*}
\textbf{Deep link}~\cite{Mobiledeeplinking} is an emerging concept in industry. The idea of deep link for mobile apps originates from the links in the Web. It is a uniform resource identifier that links to a specific page inside an app. In the context of Android mobile apps, a URI should be properly handled by configuring intent-filter, so as to bring others directly into the specific location within their app with a dedicated link. Recently, some large companies, especially search engine related, put many efforts into mobile deep linking and proposed their criteria for deep links. Google App Indexing~\cite{GoogleAppIndexing} is a system that allows people to click from listings in Google's search results into apps on their Android and iOS smartphones and tablets. Bing App Linking ~\cite{BingAppLinking} links apps to Bing search results on Windows and Windows Phone to boost the app's discoverability and engagement. Facebook App Links~\cite{FacebookAppLinks} is an open cross platform solution for deep linking to content in mobile apps. Our approach differs from these state-of-the-art solutions that these solutions require deep linked apps need to have corresponding websites to publish deep links but we don't.

Our work retrofits some prior techniques such as web crawler, automated software testing and program analysis, but is the first to combine and extend them to automate generate deep links for Android apps.

\textbf{Web Crawlers}, also known as Web spiders and (ro)bots, have been studied since the advent of the Web itself~\cite{BrinPage:Google}\cite{Burner1997}\cite{Cho:WWW2001}\cite{Heydon:1999}. There has been extensive research on the hidden Web behind forms~\cite{Barbosa:WWW2007}\cite{Dasgupta:WWW2007}\cite{Fontes:WIDM2004}. The main focus in this research area is to detect ways of accessing the Web content behind data entry points. Alvarez et al.~\cite{Alvarez:2006} discuss some challenges of crawling hidden content generated with JavaScript, but focus on hypertext links. Mesbah et al.~\cite{Mesbah:ICWE2008} was the first academic research work proposing a solution to the problem of crawling AJAX, in the form of algorithms and an open-source tool that automatically crawls and creates a finite state machine of the states and transitions. Our exploitation process brings the idea of Web crawlers, especially crawlers for the hidden Web.

\textbf{Automated Software Testing} is an effective way to ensure the quality of software. Many researchers put efforts into this research area\cite{choi2013guided}\cite{boushehrinejadmoradi2015testing}\cite{DBLP:conf/kbse/ZhangHC15}. Shauvik et al.~\cite{choudhary2015automated} presented a comparative study of the main existing test input generation techniques and corresponding tools for Android. Ravi et al.~\cite{bhoraskar2014brahmastra} presented an app automation tool called Brahmastra to the problem of third-party component integration testing at scale, in which one party wishes to test a large number of applications using the same third-party component for a potential vulnerability. Machiry et al.~\cite{machiry2013dynodroid} presented a practical system Dynodroid for generating relevant inputs to mobile apps on the dominant Android platform. It uses a novel ¡°observe-select-execute¡± principle to efficiently generate a sequence of such inputs to an app. These research work mainly focus on automated test inputs generation. Azim et al.~\cite{azim2013targeted} presented A3E, an approach and tool that allows substantial Android apps to be explored systematically while running on actual phones. The key insight of their approach is to use a static, taint-style, dataflow analysis on the app bytecode in a novel way, to construct a high-level control flow graph that captures legal transitions among activities (app screens). Our approach drew lessons from these software testing research work, and combined the test inputs generation methodology and app systematically exploration for automated deep links generation.

In industry, mature tools are developed to support automated software testing. Google android development kit provides two testing tools, Monkey~\cite{Monkey} and MonkeyRunner~\cite{Monkeyrunner}. Monkey is an automated fuzz testing tool creates random inputs without considering application¡¯s state. Hu and Neamtiu~\cite{hu2011gui} developed an useful bug finding and tracing tool based on Monkey. MonkeyRunner is a remote testing framework. A user can control application running on the phone from the computer through USB connection. There are also some other famous automated software testing tools, such as Ranorex~\cite{Ranorex} and Robotium~\cite{Robotium}. We adopt Monkeyrunner to stimulate click operation on Android phones. In order to obtain the content from a webview-based app, we adopted xposed~\cite{xposed}, a framework for modules that can change the behavior of the system and apps.

\textbf{Static Program Analysis}, also known as code analysis, is the analysis of software that is performed without actually executing programs. Paulo et al.~\cite{barros2015static} presented static analyses for two types of implicit control flow that frequently appear in Android apps: Java reflection and Android intents. Their analyses help to resolve where control flows and what data is passed. Bastani et al.\cite{bastani2015interactively} proposed a process for producing apps certified to be free of malicious explicit information flows. In their approach, the developer provides tests that specify what code is reachable, allowing the static analysis to restrict its search to tested code. More recently, Damien et al.~\cite{octeau2016combining} shows how to overlay a probabilistic model, trained using domain knowledge, on top of static analysis results, in order to triage static analysis results. They apply this idea to analyzing mobile applications. Android application components can communicate with each other, both within single applications and between different applications. In our approach, we apply static code analysis to implement the Navigation Analyzer, which analyzes navigation among activities and builds an Navigation Graph.

%% file: Conclusion.tex
In this paper, we have presented an approach to automatically generating deep links for mobile apps on Android. We first designed a model that is suitable for automatic generation of deep links of the content page in Android apps. Guided by the model, our approach consisted of three processes: (1) \textbf{exploration} process that identifies how pages inside apps can be reached with least transitions; (2) \textbf{exploitation} process that publishes deep links of apps to make them discoverable and accessible to third parties; and (3) \textbf{execution} process that handles the actual query and invocation of deep links from third-party apps or services. We evaluated the feasibility and efficiency of our approach on popular Android apps.

We are now deploying our deep link model in a top Android app stores in China\footnote{The exact name is hidden here due to double-blind review requirements. We will release the details when the manuscript is published}. Developers can easily repackage their apps by our toolkits to equip the deep links. We also incorporate the back-end cloud supports of this app store to maintain the deep link and realize app indexing, so that the users and other app developers can search, browse, and consume the data of apps that comply with our model and are published on the app store.